\documentclass[aps,prx,twocolumn,superscriptaddress,amssymb,showpacs,notitlepage]{revtex4-2}
\usepackage{bm}
\usepackage{graphicx,hyperref}
\usepackage{dcolumn}
\usepackage{braket}
\usepackage{natbib}
\usepackage{amsmath}
\usepackage{mathptmx}
\usepackage{color}
\usepackage{siunitx}
\usepackage{mathtools}
\usepackage{soul}
\usepackage{float}
\usepackage{amssymb}
\usepackage{esint}
\usepackage{physics}
\usepackage{makecell}
\usepackage{multirow}
\usepackage{array}
\usepackage{url}
\usepackage{adjustbox}
\usepackage{minted}
\usepackage{listings}
\usepackage{threeparttable,booktabs}
\usepackage{fontawesome}
\usepackage{cleveref}
\usepackage{svg}
\usepackage[normalem]{ulem}

\begin{document}
\title{Probing Quantum Anomalous Hall States in Twisted Bilayer WSe$_2$ via Attractive Polaron Spectroscopy}

\author{Beini Gao}
\affiliation{Joint Quantum Institute (JQI), University of Maryland, College Park, MD 20742, USA}
\author{Mahdi Ghafariasl}
\altaffiliation{These authors contributed equally to this work}
\affiliation{Joint Quantum Institute (JQI), University of Maryland, College Park, MD 20742, USA}
\author{Mahmoud Jalali Mehrabad}
\altaffiliation{These authors contributed equally to this work}
\affiliation{Joint Quantum Institute (JQI), University of Maryland, College Park, MD 20742, USA}
\author{Tsung-Sheng Huang}
\altaffiliation{These authors contributed equally to this work}
\affiliation{Joint Quantum Institute (JQI), University of Maryland, College Park, MD 20742, USA}
\author{Lifu Zhang}
\affiliation{Joint Quantum Institute (JQI), University of Maryland, College Park, MD 20742, USA}
\author{Deric Session}
\affiliation{Joint Quantum Institute (JQI), University of Maryland, College Park, MD 20742, USA}
\author{Pranshoo Upadhyay}
\affiliation{Joint Quantum Institute (JQI), University of Maryland, College Park, MD 20742, USA}
\author{Rundong Ma}
\affiliation{Electrical Engineering Department, University of Maryland, College Park, MD 20742, USA}
\author{Ghadah Alshalan}
\affiliation{Joint Quantum Institute (JQI), University of Maryland, College Park, MD 20742, USA}
\author{Daniel Suarez}
\affiliation{Joint Quantum Institute (JQI), University of Maryland, College Park, MD 20742, USA}
\author{Supratik Sarkar}
\affiliation{Joint Quantum Institute (JQI), University of Maryland, College Park, MD 20742, USA}
\author{Suji Park}
\affiliation{Center for Functional Nanomaterials, Brookhaven National Laboratory, Upton, NY 11973, USA.}
\author{Houk Jang}
\affiliation{Center for Functional Nanomaterials, Brookhaven National Laboratory, Upton, NY 11973, USA.}
\author{Kenji Watanabe}
\affiliation{Research Center for Electronic and Optical Materials, National Institute for Materials Science, 1-1 Namiki, Tsukuba 305-0044, Japan.}
\author{Takashi Taniguchi}
\affiliation{Research Center for Electronic and Optical Materials, National Institute for Materials Science, 1-1 Namiki, Tsukuba 305-0044, Japan.}
\author{Ming Xie}
\affiliation{Condensed Matter Theory Center, University of Maryland, College Park, MD, USA}
\author{You Zhou}
\thanks{Corresponding author: youzhou@umd.edu}
\affiliation{Department of Materials Science and Engineering, University of Maryland, College Park, MD 20742, USA}
\author{Mohammad Hafezi}
\thanks{Corresponding author: hafezi@umd.edu}
\affiliation{Joint Quantum Institute (JQI), University of Maryland, College Park, MD 20742, USA}  

\date{\today}
\begin{abstract}

Moir\'e superlattices in semiconductors exhibit a rich variety of interaction-induced topological states, including quantum anomalous Hall (QAH) effects~\cite{cai2023signatures,zeng2023thermodynamic,park2023observation,xu2023observation,ji2024local,redekop2024direct,xu2025interplay,park2025ferromagnetism}. A recent study hinted that twisted WSe\(_2\) homobilayer (tWSe$_2$) could host a QAH state but lacked direct evidence of ferromagnetism, a key hallmark of this phase~\cite{foutty2024mapping}. Here, we report the first direct evidence of QAH states in tWSe$_2$ with spontaneous ferromagnetism.  Specifically, we employ polarization-resolved attractive polaron spectroscopy on a dual-gated, 2\(^\circ\) tWSe$_2$ and observe direct signatures of spontaneous time-reversal symmetry breaking at hole filling \(\nu = 1\).  Together with a Chern number ($C$) measurement via Streda formula analysis, we identify this magnetized state as a topological state, characterized by \(C = 1\). Furthermore, we demonstrate that these topological and magnetic properties are tunable via a finite displacement field, between a QAH ferromagnetic state and an antiferromagnetic state. Our findings position tWSe\(_2\) as a highly versatile, stable, and optically addressable platform for investigating topological order and strong correlations in two-dimensional landscapes.

\end{abstract}

\maketitle
\begin{center}
\textbf{\large I. Introduction}\\
\end{center}

Twisted transition metal dichalcogenide (TMD) bilayers have emerged as a highly tunable platform for investigating correlated electronic phases in two dimensions. 
In these systems, the presence of periodic moiré potentials --- arising from lattice mismatch --- produces strong interaction and non-trivial topology, resulting in a wide array of exotic many-body states.  These include superconductivity~\cite{xia2025superconductivity,guo2025superconductivity}, ferroelectric order~\cite{weston2022interfacial}, Mott insulators~\cite{Regan2020,tang2020simulation,xiong2023correlated,gao2024excitonic}, generalized Wigner crystals~\cite{Regan2020, xu2020correlated, zhou2021bilayer}, and a growing set of topological phases, such as integer and fractional quantum spin Hall~\cite{kang2024evidence} and quantum anomalous Hall (QAH) states~\cite{cai2023signatures,zeng2023thermodynamic,park2023observation,xu2023observation,ji2024local,redekop2024direct,xu2025interplay,park2025ferromagnetism}.

QAH phases are especially appealing as they exhibit topological properties even in the absence of an external magnetic field. Examples include integer and fractional QAH states in MoTe$_2$~\cite{cai2023signatures,zeng2023thermodynamic,park2023observation,xu2023observation,park2025ferromagnetism,ji2024local,redekop2024direct,xu2025interplay}, and integer QAH in twisted MoTe$_2$–WSe$_2$ heterobilayers~\cite{li2021quantum}. Despite these advances, tWSe$_2$ remains largely underexplored even though it offers practical advantages, including enhanced air stability and visible-frequency optical transitions --- features that make it an excellent platform for scalable device engineering and optical interrogation. Recent breakthrough experiments in tWSe$_2$ have demonstrated QAH effect at 1.2° twist angle~\cite{foutty2024mapping}, as well as superconductivity at 3.6° and 5° angles~\cite{xia2025superconductivity,guo2025superconductivity}. However, ferromagnetism in the QAH states has yet to be unambiguously observed. The precise twist angular dependence of these emergent states and their potential interplay has become a topic of significant interest.
\begin{figure*}
    \centering    
    \includegraphics[width=\textwidth]{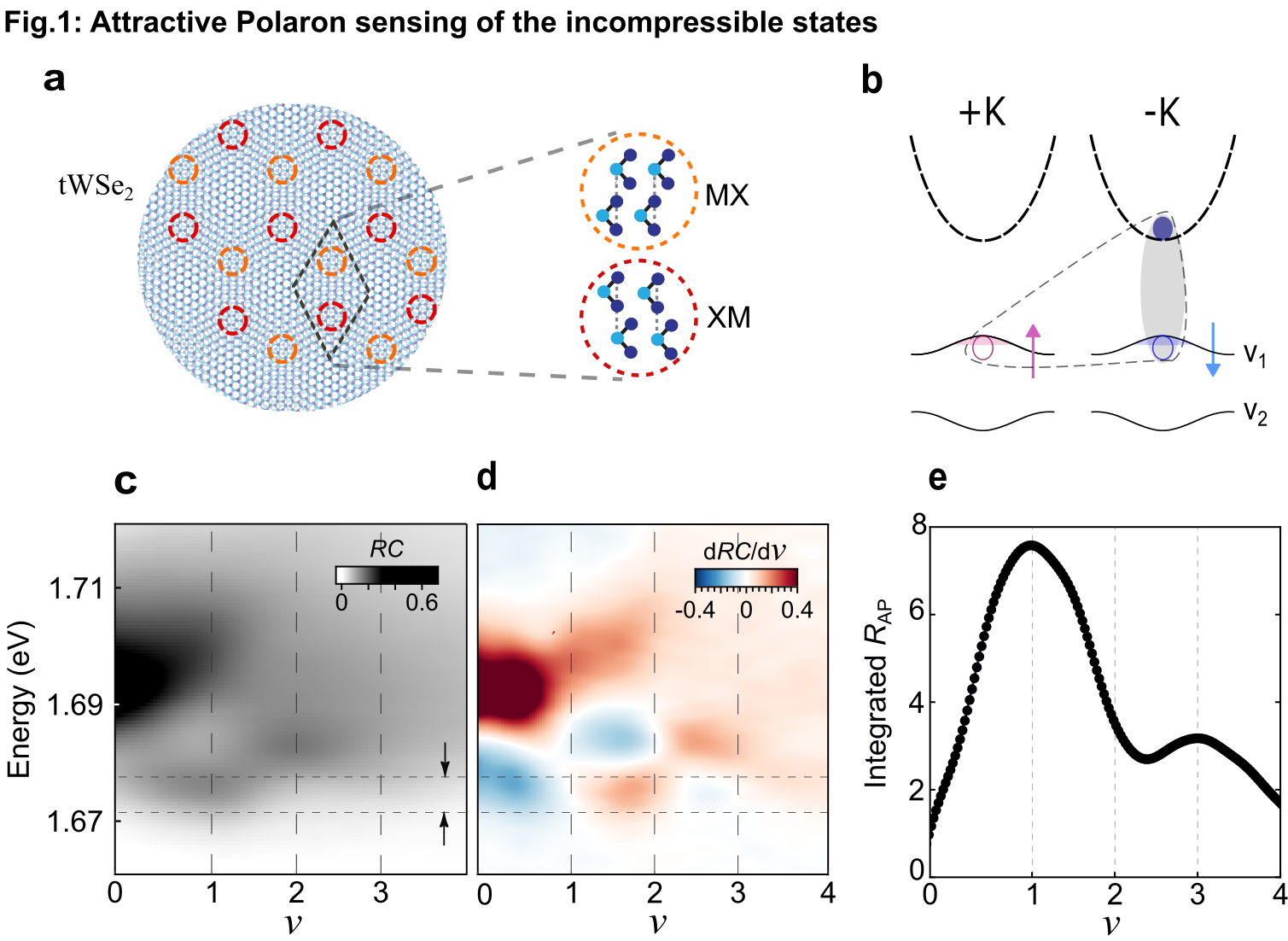}
    \caption{ 
   \textbf{a,} Illustration of the moiré pattern in a R-stacked twisted $\text{WSe}_2$ homobilayer. Orange and red circles denote the high-symmetry atomic registries MX and XM, respectively (see inset).
    \textbf{b,} Schematic of an attractive polaron (AP) formed by an exciton in one valley (here at $-K$) and a hole in the opposite valley ($+K$).
    The exciton consists of an electron in the lowest conduction moiré band (dashed) and a hole in the topmost valence moiré band ($V_1$). A second valence moir\'e band, $V_2$, is separated from $V_1$.
   \textbf{c,} Reflection contrast (RC) spectrum, defined as $\frac{R-R_0}{R_0}$, at temperature 4K. The hole-filling fraction $\nu$ is tuned by gates (see Methods for calibration). The horizontal dashed lines (and arrows) indicate the integrated energy range of the attractive polaron. 
   \textbf{d,} Derivative of RC with respect to $\nu$.  
   \textbf{e,} Doping dependence of the integrated AP intensity ($R_{\mathrm{AP}}$).  
   }
    \label{fig1}
\end{figure*}

Specifically, the magnetic and topological nature of correlated states in this system remains largely uncharted. For instance, clear signatures of spontaneous time-reversal symmetry breaking at integer fillings, and probe of the phase diagram in tWSe$_2$ have yet to be observed. 

Here, we report the first optical detection of ferromagnetic and the quantum anomalous Hall state in tWSe$_2$, using attractive polaron (AP) spectroscopy~\cite{ciorciaro2023kinetic}. Polarization-resolved reflection near the AP resonance reveals clear signatures of spontaneous magnetization at hole filling $\nu = 1$. Using the Streda formula, we identify this state as a Chern insulator with a Chern number $C = 1$,  intriguingly opposite in sign to that reported in an earlier study in 1.23 $^\circ$ tWSe$_2$ ~\cite{foutty2024mapping}. Our optical measurement establishes the presence of topological states on a macroscopic level, which is inaccessible via local scanning probes~\cite{foutty2024mapping}.  Furthermore, we demonstrate electric field control of a QAH ferromagnetic phase to an antiferromagnetic state in such a system. 
\\

\begin{center}
\textbf{\large II. Attractive Polaron Spectroscopy}\\
\end{center}

Our sample is 2° twisted WSe$_2$ with a hexagonal structure whose superlattice periodicity is approximately 9.4 $\rm{nm}$ (see Supplemental Information Fig. 2 for calibration of twisting angle). The two sublattices in a moir\'e unit cell are denoted by MX and XM, respectively, as shown in Fig.~\ref{fig1}a (see Supplemental Information Fig. 1 for  the calculated electronic structure). 
We employ a dual-gate geometry to enable independent control of both the hole filling factor, $\nu$, and the perpendicular displacement field ($D$). Tuning $D$ controls the interlayer potential difference in tWSe$_2$.

The attractive polaron in our system is a composite excitation between an exciton in one valley and a hole in the opposite valley, as shown schematically in Fig.~\ref{fig1}b.
Therefore, the polarization-resolved optical response at AP resonance reveals the hole occupancy in the opposite valley \cite{ciorciaro2023kinetic}.

\begin{figure*}
    \centering    
    \includegraphics[width=\textwidth]{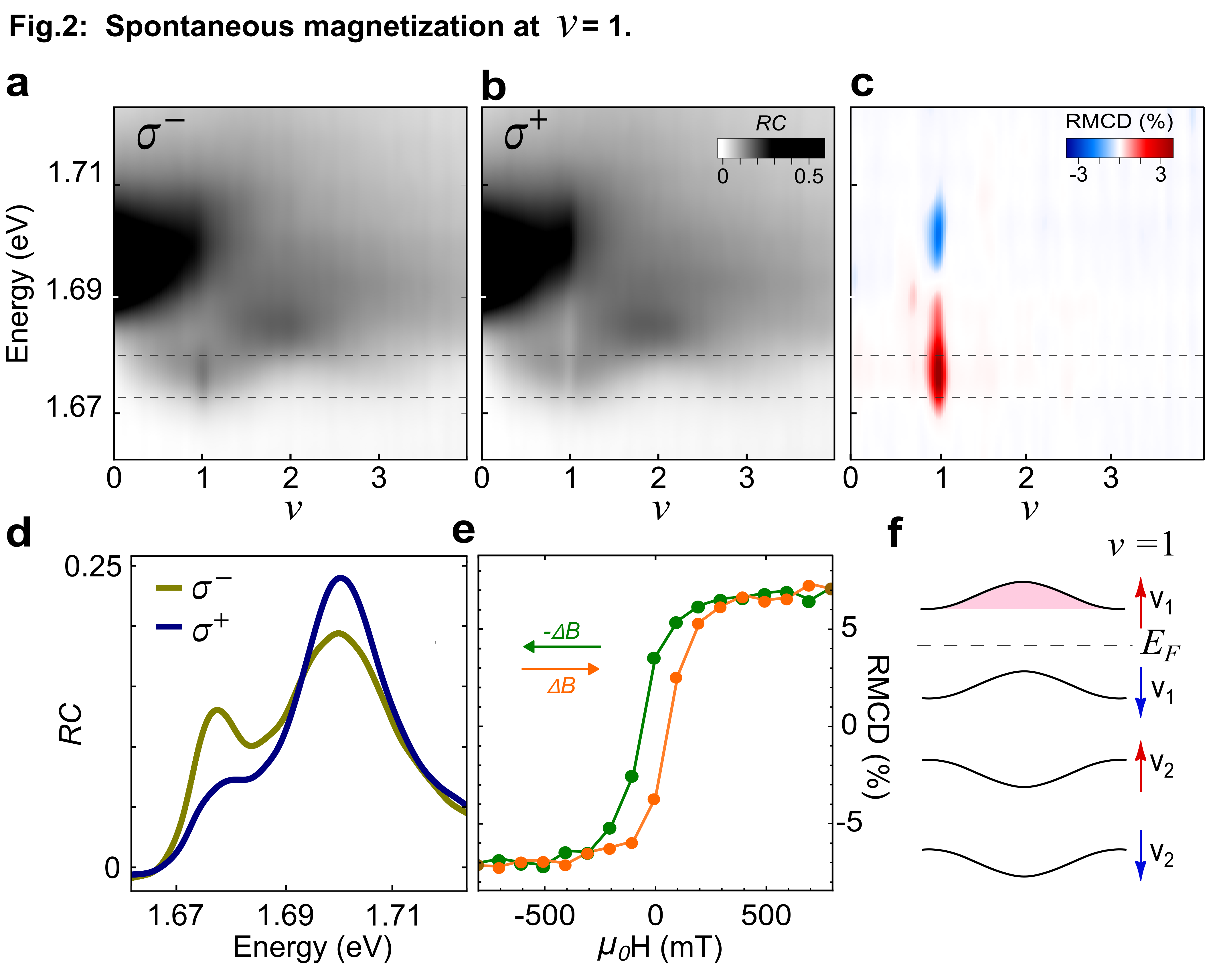}
    \caption{  
    RC as a function of energy and hole filling fraction $\nu$ under    \textbf{a,} $\sigma^-$ and    \textbf{b,} $\sigma^+$ excitation, respectively.  
   \textbf{c,} Spectrally resolved reflection magnetic circular dichroism (RMCD) at various $\nu$.  
   \textbf{d,} Polarization-resolved reflection spectrum at $\nu = 1$.   
   \textbf{e,} Magnetic hysteresis loop measured at $\nu = 1$; green and orange indicate the scan directions with respect to the magnetic field ($\pm\Delta B$). \textbf{f,} Schematic of the \textit{many-body} bands at $\nu = 1$, incorporating electrostatic interaction at the mean-field level~\cite{wu2019topological}. v$_1$ and v$_2$ denote the first and the second moiré valence bands, respectively. This diagram is specific to $\nu=1$ and does not represent the band structure at other fillings.} All the data were taken at a lattice temperature $T_{\rm{lattice}}=66$ mK.
    \label{fig2}
\end{figure*}

To identify and characterize the AP resonance, we measure the reflection contrast ($RC$) spectrum as a function of $\nu$, which exhibits several resonances (Fig.~\ref{fig1}c). Near the charge-neutral regime, we identify tWSe$_2$ intralayer excitons at 1.694 eV. As the system is hole-doped, we observe a continuously blue-shifting resonance of the repulsive polaron, and a red-shifting resonance (upon doping until $\nu =1$) as the AP. We note that oscillator strength is transferred to a higher energy intrinsic species near $\nu = 2$. By analyzing the doping-dependent derivative of $RC$, we identify two sign reversals at $\nu=1$ and $\nu=3$ (Fig.~\ref{fig1}d), marking the peak of the oscillator strength of the AP. By integrating the AP intensity (within the horizontal dashed lines indicated in Fig.~\ref{fig1}c), we observe that its oscillator strength increases monotonically with $\nu$ up to $\nu=1$, then decreases between $\nu=1$ and $\nu=2$, as shown in Fig.~\ref{fig1}e  indicating a Hubbard picture where the AP oscillator strength scales with the number of singly occupied states in the valence moiré bands \cite{wang2020correlated,pan2020band,zang2021hartree,munoz2025twist} (see Supplemental Information Fig. 1). The same trend repeats between $\nu=2$ and $\nu=4$.

To qualitatively understand the observed characteristics of AP, we use the fact that its oscillator strength scale with the number of singly occupied states in the valence moir\'e bands \cite{ciorciaro2023kinetic}. In other words, any creation of vacant and doubly occupied states transfers the AP oscillator strength to other species outside of the AP's integrated frequency range. Here, we note that at $\nu=1$, there is a single hole per moir\'e unit cell, and therefore, gives the maximum number of singly occupied states for the topmost moir\'e valence band v$_1$ (Fig.~\ref{fig1}b). The observed repeating oscillator strength trend for $\nu>2$ in Fig.~\ref{fig1}e follows a similar explanation for the second moir\'e band v$_2$. In addition, independent measurements using a MoSe$_2$ sensing layer (see Supplemental Information Fig. 16) reveals suppressed compressibility at $\nu=1$, corroborating that the AP maximum corresponds to a correlated insulating state.
\\

\begin{figure*}
    \centering    
    \includegraphics[width=\textwidth]{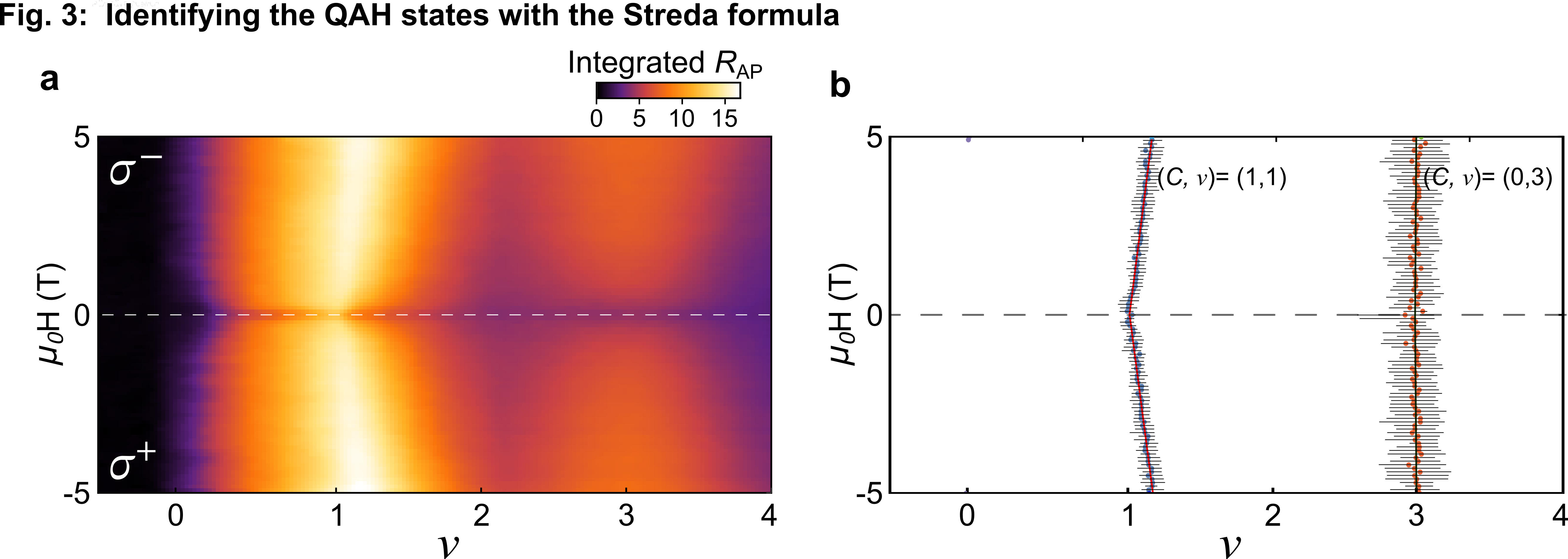}
    \caption{
    \textbf{a,} Integrated AP intensity as a function of magnetic field $\mu_0H$ and hole filling $\nu$.  
    \textbf{b,} Extracted maxima of the integrated AP intensity at each $\mu_0H$ near $\nu \simeq 1, 3$, indicated by blue, and orange points, respectively. Error bars represent 3\% deviation from the centroid for $\nu=1$ and 5\% deviation for $\nu=3$ (see Methods).
    Red and dark green lines near $\nu = 1$ and $\nu = 3$ indicate a linear fit to the data near $\nu=1$ and $\nu=3$, from which the slope indicates Chern number $C=1$ for $\nu = 1$, and $C=0$ for $\nu = 3$.}
    \label{fig3}
\end{figure*}
\begin{center}
\textbf{\large III. Spontaneous magnetization at $\nu = 1$}\\
\end{center} 

To optically detect spontaneous magnetization of the system, we measure polarization-resolved $RC$ under zero magnetic field (Fig.~\ref{fig2}a-b) and construct the reflection magnetic circular dichroism (RMCD) as a function of $\nu$ at 66 mK, as shown in Fig.~\ref{fig2}c. Since the RC from $\sigma_-$ and $\sigma_+$ excitations selectively probe the occupation of doped charges in the $+K$ and $-K$ valleys, respectively, any valley and spin imbalance will lead to a different optical response in the two polarizations. Remarkably, we observe that the $RC$ exhibits a pronounced peak for $\sigma_-$ excitation at $\nu = 1$, while a dip is observed for $\sigma_+$ within the AP energy range, a direct indication of spontaneous time-reversal symmetry breaking.  To be more quantitative, the RMCD measurement in Fig.~\ref{fig2}c captures the out-of-plane magnetization of the doped charges. The magnetization peaks at $\nu = 1$ (see Fig.~\ref{fig2}d for $RC$ spectra), while it remains negligible at other fillings.

Crucially, in addition to spontaneous RMCD under a zero field, we observe a magnetic hysteresis loop at $\nu = 1$, which indicates the presence of ferromagnetism (FM), with a coercive magnetic field of approximately 50 mT, as shown in Fig.~\ref{fig2}e. The observed emergent magnetism can be understood in the mean-field approximation, where electron-electron interactions generate an effective exchange splitting between the two valleys, yielding FM around $\nu = 1$ as only one valley is occupied, as schematically illustrated in Fig.~\ref{fig2}f. 
The hysteresis disappears at a higher temperature of around 1.2 K, see Supplemental Information Fig. 3.
Notably, our observation of FM at $\nu = 1$ in tWSe$_2$ is similar to the general trend seen in MoTe$_2$; however, here we report the absence of FM at $\nu=3$, in contrast to MoTe$_2$.
\\

\begin{center}
\textbf{\large IV. Identifying the quantum anomalous Hall state with Streda formula}\\
\end{center}

To directly determine the topological properties of our system, we measure the magnetic field- and doping-dependence of the optical spectra and extract the topological invariant from the Streda formula \cite{cai2023signatures,xie2021fractional,nuckolls2020strongly,spanton2018observation}. Specifically, we track the frequency-integrated oscillator strength, as a function of magnetic field $\mu_0H$ and gate voltage (see Supplemental Information Fig. 4-6 for details). In particular, the AP spectroscopy performed at $\nu>0$ allows us to track the density shift of correlated states as a function of the magnetic field $\mu_0H$, as shown in Fig.~\ref{fig3}a.
The local maxima therein for each $\mu_0H$ is extracted and shown in Fig.~\ref{fig3}b, and the slope of each line indicates the Chern number ($C$) of the corresponding state.

The polaron peak at $\nu=1$ marks a correlated insulating state whose density shifts linearly as a function of the magnetic field, with a slope that matches $C=1$, according to the Streda formula. The sign of the slope indicates that the total magnetization of the $\nu=1$ ground state has the same sign as the Chern number. Interestingly, this measured Chern number has the opposite sign compared to recent reports in tWSe$_2$
 ~\cite{foutty2024mapping} with a smaller twist angle of 1.23$^\circ$, potentially due to the twist angle difference and lattice relaxation ~\cite{zhang2024polarization,devakul2021magic}. Together with the emergence of FM at this filling, we attribute this state to an integer QAH phase. Upon reversal of the magnetic field, the magnetization reverses, which also leads to a reversal of the Chern number to $C=-1$, manifesting a valley contrasting Chern number for the many-body ground state. Compared to twisted MoTe$_2$, the lack of an experimentally observed QAH phase in earlier experiments employing twisted WSe$_2$ ~\cite{knuppel2025correlated} may be attributed to its significantly lower ferromagnetic transition temperature, $T_C <$ 1.2 $K$, in contrast to $T_C \approx$ 10 $K$ in MoTe$_2$. We also note that at twist angles smaller than the current studies, such as $\theta \leq 1.5^{\circ}$, tWSe$_2$ can suffer a stronger lattice relaxation effect compared to tMoTe2. This can lead to high twist angle disorder, as was the case in earlier local probe measurement ~\cite{foutty2024mapping}.

In contrast to the QAH state at $\nu=1$ , the feature at $\nu = 3$ shows vanishing Streda slope and therefore is compatible with a zero Chern number $C = 0$. This may suggest a topologically trivial correlated phase, such as an intervalley coherent or antiferromagnetic state, which has not been observed previously reported in higher moiré bands of this system \cite{park2025ferromagnetism,kang2024evidence,kang2024double}, or potentially a non-Abelian spin Hall insulator, as theoretically predicted \cite{abouelkomsan2025non}.Further investigation will be required to clarify  teh character of the $\nu=3$ state.
\\

\begin{center}
\textbf{\large V. Displacement field tuning: from QAH FM to AFM}\\
\end{center}
\begin{figure*}
    \centering    
    \includegraphics[width=\textwidth]{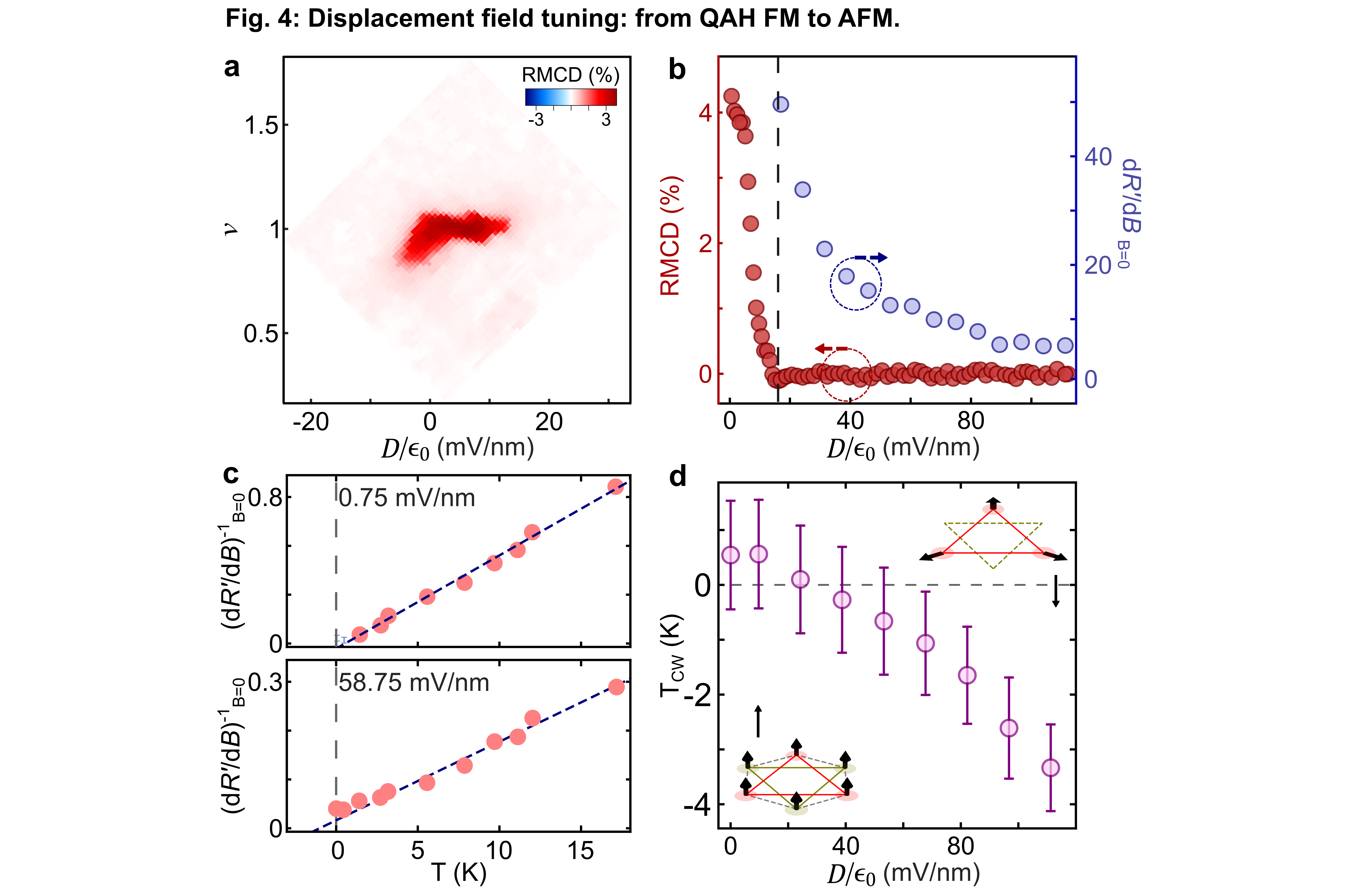}
    \caption{
    \textbf{a,} RMCD as a function of the displacement field $D$ and hole filling fraction $\nu$ at magnetic field $B=0$ and temperature $T_{\rm{lattice}}$=66 mK. \textbf{b,} (Red, left y-axis) Reflection magnetic circular dichroism (RMCD) as a function of the displacement field $\frac{D}{\epsilon_0}$ at $B=0$. (Blue, right y-axis) derivative of RMCD with respect to magnetic field $B$ at $B=0$ measured as a function of $\frac{D}{\epsilon_0}$.
    \textbf{c, } Magnetic susceptibility as a function of temperature for two displacement field values. The dashed line is the Curie-Weiss fit. All the small displacement field fittings are done with data points above the Curie temperature of $\sim$1.2 K, see Supplemental Information Fig. 3, Fig. 9.
    \textbf{d, } Curie-Weiss temperature $T_{CW}$ as a function of $\frac{D}{\epsilon_0}$. Linear fitting error bars are also shown. The inset schematics indicate the (bottom) QAH FM phase and the (top) AFM phase, respectively. }
    \label{fig4}
\end{figure*}

With QAH physics at $\nu=1$ established under a zero displacement field, we further investigate the robustness and tunability of this Chern insulating state with a finite displacement field $D/\epsilon_0$.
First, in Fig.~\ref{fig4}a, we present the RMCD diagram with respect to $\nu $ and $D/\epsilon_0 $at zero magnetic field and lattice temperature of 66 mK. We observe a prominent RMCD response at around $\nu=1$, where the RMCD signal persists up to $\sim 18$ mV/nm. This region of a prominent RMCD response establishes the ferromagnetic regime in the overall phase diagram of tWSe$_2$. At a slightly elevated magnetic field of 0.1 T, the ferromagnetic phase becomes further stabilized across a broader regime (Supplemental Information Fig. 8). Notably, ferromagnetic order is stabilized near the van Hove singularity  , suggesting enhanced interaction strength due to the high density of states(Supplemental Information Fig. 8) ~\cite{guo2025superconductivity,xia2025superconductivity,knuppel2025correlated}.

Focusing on the displacement-field-dependence of the RMCD at $\nu = 1$, we observe that the magnetization of the system quenches above a critical displacement field of 18 mV/nm, as indicated by the dashed line in Fig.~\ref{fig4}b. Meanwhile, near this critical displacement field, we also observe a divergent magnetic susceptibility ($\chi=dR'/dB$, where $R'$ stands for RMCD), consistent with a field-controlled magnetic phase transition (Fig.~\ref{fig4}b, see Supplemental Information Fig. 7 for spectrally resolved data).

To understand the origin of this magnetic phase transition, we measure the magnetic susceptibility of the sample by performing temperature-dependent RMCD measurements and extract the Curie-Weiss temperature $T_{CW}$ (see Supplemental Information Fig. 9 for details). Intriguingly, we observe a clear sign-change of the Curie-Weiss temperature $T_{CW}$, as shown in Fig.~\ref{fig4}c, from positive to negative with increasing $D$. This indicates a transition from FM to antiferromagnetic (AFM) order~\cite{pan2020band,anderson2023programming} in our system, due to a change in the lattice geometry as shown schematically in the inset of Fig.~\ref{fig4}d. A finite $D$ introduces a layer-dependent potential that polarizes the moiré valence band toward one physical layer, thereby suppressing interlayer hybridization and redistributing the Berry curvature. Our numerical calculations (Supplemental Information Fig. 10) confirm that increasing interlayer bias progressively shifts the Berry curvature weight from the $K$ region toward $K'$, driving the transition from $C=1$ to $C=0$. Together, these findings reveal a field-tunable magnetic phase transition linked to the underlying topological properties of the correlated moir\'e system. 
\\

\begin{center}
\textbf{\large Conclusion}\\
\end{center}
In summary, we have reported the first optical detection of the QAH states in a tWSe\(_2\) system and demonstrated displacement-field tunability of the underlying magnetic order, showcasing transitions between quantum anomalous Hall ferromagnetic and antiferromagnetic phases. These results highlight tWSe\(_2\) as a uniquely versatile platform, combining visible-frequency optical access with air stability and fabrication simplicity—features that open new directions for probing correlated and topological quantum phases in moiré systems.

Looking ahead, several exciting research avenues can be built upon our findings. One promising direction is the exploration of fractional Chern insulators~\cite{crepel2023anomalous}, which have recently been observed in tMoTe\(_2\)~\cite{cai2023signatures,zeng2023thermodynamic,park2023observation,xu2023observation,chang2025emergentchargetransferferromagnetismfractional,wang2025hiddenstatesdynamicsfractional,anderson2025magnetoelectriccontrolhelicallight}, and whose realization in tWSe$_2$ would open access to new correlated topological phases. 
Furthermore, similar to the recently demonstrated coexistence of QAH and superconductivity in tMoTe$_2$ \cite{xu2025signaturesunconventionalsuperconductivitynear}, we anticipate intriguing interplay between topological states and superconductivity in tWSe$_2$\cite{guo2025superconductivity,xia2025superconductivity}.
Moreover, the presence of different Chern numbers in higher moire bands compared to MoTe$_2$ may lead to the realization of other correlated states ~\cite{zhang2024polarization}. 
More intriguing is the realization of recently predicted topological excitons~\cite{xie2024long}, accessible in a modified device architecture designed to host long-lived interlayer excitons. The coexistence of strong exciton-exciton interactions~\cite{park2023dipole,gao2024excitonic,xiong2023correlated,lian2024valley} and tunable excitonic band topology~\cite{xie2024long} offers a compelling pathway toward the realization of bosonic fractional Chern insulators~\cite{hafezi2007fractional,moller2009composite, wang2011fractional}. Such long-sought phase are bosonic analogue of fermionic fractional Chern insulators~\cite{regnault2011fractional,neupert2011fractional, sheng2011fractional,tang2011high,sun2011nearly},  which has remained unobserved in any system to date. More generally, this system could be considered as a promising candidate for optical probe and manipulation of strongly-correlated electron-photon systems~\cite{bloch2022strongly, basov2011electrodynamics}. Specifically, interesting prospects are optical detection and manipulation of topological states~\cite{mcginley2024signatures,nambiar2024diagnosing}. 

\begin{center}
\textbf{\large Methods}\\
\end{center}
\begin{center}
\textbf{Device fabrication}\\
\end{center}
The device was made by ``tear-and-stack'' technique using the standard polymer-based dry transfer method. With picking up the top gate graphite, top gate hBN, tWSe$_2$, bilayer hBN spacer, half overlapped MoSe$_2$ monolayer in sequence and dropped down onto a pre-prepared hBN-Graphite bottom gate on a 285 nm SiO$_2$/Si substrate. Followed by electron beam lithography and metal deposition for the gate and contacts (5 nm Cr, 70 nm Au). All the monolayers are identified via optical contrast. The top ($\sim$ 17.8 nm) and bottom ($\sim$ 24.6 nm) dielectric hBN thickness are confirmed by atomic force microscopy.  
\\
\begin{center}
\textbf{Optical measurements}\\
\end{center}

Optical measurements were performed using confocal microscopy in an optically accessible dilution fridge (BlueFors) with vertical magnetic fields up to 9T and temperatures down to a lattice temperature of 60 mK~\cite{Jonathan2022thesis}. A halogen lamp was used as the light source. The input light was spectrally filtered to 650 to 900 nm range. A low-temperature microscope objective was used to focus the light onto the sample. The light intensity on the sample was kept around 0.14 nW/$\mu$m$^2$ to minimize its effects on the electronic states. 
The reflected light was collected by the same objective and analyzed by a spectrometer equipped with a charge coupled device array to obtain spectrum. The RC spectrum is defined as $(R-R_0)/R$, where $R_0$ the reference spectrum was taken for the sample at a high doping density with quenched excitonic resonances.
For all the zero magnetic field data presented in Fig.~\ref{fig2} and ~\ref{fig4}, we polarize the system with a positive vertical magnetic field and then ramp back to perform the zero magnetic field measurement.
The RMCD was used to study the magnetic properties of the samples. A combination of a linear polarizer and a quarter-wave plate was used to generate circularly polarized light ($\sigma_-$ and $\sigma_+$) on the sample. The RMCD spectrum is defined as $\frac{\sigma_--\sigma_+}{{\sigma_-+\sigma_+}}$.

For the AP spectroscopy, we integrate the reflectance over a range of wavelengths (738-741.2 nm, or equivalently, 1.673 - 1.680 eV) that focuses on the attractive polaron resonance of tWSe$_2$. The same wavelength range was used to obtain the average of the absolute value of RMCD. The magnetic susceptibility was evaluated from the slope of MCD at zero magnetic field.

For the streda formula data presented in Fig.~\ref{fig3}, we use $RC$ under $\sigma^-$ excitation for the positive magnetic field detection, and $\sigma^+$ excitation for the negative magnetic field detection.
\\

\begin{center}
\textbf{Twist Angle Calibration}\\
\end{center}

We calibrate the twist angle of the tWSe$_2$ presented in the main text using the quantum oscillations observed optically under an out-of-plane magnetic field of 8.8T. In Supplemental Information Fig. 2, we present the doping dependent $RC$ response of a monolayer tWSe$_2$ under $\sigma_+$ excitation.
The oscillation signal formed as a result of the formation of the Landau levels (LLs). The LL period is determined to be 0.244 V for the single gate response, from which we deduce a hole density change of 1.75$\times 10^{12}$ cm$^{-2}$ per volt for data presented in the main text. Combining with the moir\'e period $\Delta V_M$ extracted from the doping dependent $RC$ data, we obtain a moir\'e density $n_M$ of 1.31$\times 10^{12}$ cm$^{-2}$. 
The twist angle calibration is further corroborated by the doping density calibration, as described in the next section.
\\
\begin{center}
\textbf{Doping density calibration}\\
\end{center}

A parallel-plate capacitor model is used to calculate the gate-induced carrier density $n$ and displacement field $D$ based on the applied gate voltages. The gate capacitances per area of the top and bottom gates, $C_{t/b}=\epsilon_{hBN}\epsilon_0/d_{t/b}$, are determined by the hBN thickness $d_{t/b}$ and dielectric constant. We use $\epsilon_{hBN}=$ 3.89 here. The sample doping density $n$ can be calculated as $n=(C_t V_t+C_b V_b)$, where $V_t$ and $V_b$ are the voltages applied to the top and bottom gates. The displacement field is defined as $D=(C_t V_t-C_b V_b)/2\epsilon_0-D_{\rm{offset}}$, where $\epsilon_0$ is the vacuum permittivity and the subtracted value is the built-in offset. \\
\begin{center}
\textbf{Extraction and analysis of the Streda signal}\\
\end{center}

To extract the Streda response, we first obtain polarization-resolved reflection-contrast (RC) spectra as a function of gate voltage and magnetic field. For each magnetic-field value, an energy window containing the attractive-polaron (AP) resonance is integrated to yield an intensity map $I_{\rm{AP}}(\nu ,\mu_0 H)$, where $\nu$ is the hole filling. The resulting $I_{\rm{AP}}(\nu)$ curve exhibits local maxima at incompressible fillings. We identify this maximum from the derivative of $I_{\rm{AP}}$ with respect to $\nu$ and treat its position $\nu_{\rm{max}}(B)$ as the quantity entering the Streda analysis.

The uncertainty in each extracted $\nu_{\rm{max}}$ is estimated by computing the centroid of the AP peak at fixed magnetic field and defining an error range corresponding to a deviation of the integrated intensity from the centroid. For the sharp $\nu=1$ feature we use a 3\% deviation; for the broader 
$\nu=3$ feature we use 5\%.

The Streda slope $d\nu/dB$ is then obtained from a linear fit to $\nu_{\rm{max}}(B)$ near the filling of interest. Using the Streda relation $C=\frac{h}{e} \frac{d\nu}{dB} $, we extract the corresponding Chern number. To verify robustness, we repeat the analysis using several different integration windows for the AP resonance (widths of 3.05 meV, 5.71 meV, and 8.85 meV). All choices yield consistent slopes within the extracted uncertainties. \\

\textbf{Acknowledgements:}
We thank Ataç İmamoğlu, Liang Fu, Sankar Das Sarma, Ajit Srivastava, Patrick Kn\"{u}ppel, Xiaowei Zhang for fruitful discussions.
\\

\textbf{Author Contributions:} 
B.G., Y.Z., and M.H. conceived the project. B.G. performed the optical measurements and analyzed the data. M.X. did the theoretical calculation. M.X. and T.S.H. contributed to the theoretical understanding. L.Z. and B.G. fabricated the device. L.Z., M.G., D.S., P.U., G.A., R.M., M.J.M, and S.S. assisted with the experiment. S.J. and H.J. exfoliated the TMD flakes. T.T. and K.W. provided the bulk h-BN crystals. All authors contributed the writing of the manuscript. Y.Z. and M.H. supervised the work. All authors discussed the results and contributed to the manuscript.
\\

\textbf{Competing interests:} The authors declare no competing interests.
\\

\textbf{Data availability:}
All of the data that support the findings of this study are reported in the main text and Supplementary Information. Source data are available from the corresponding authors on reasonable request.

\bibliography{biblio.bib}

\end{document}